# MECHANISM AND STATISTIC THEORY OF
# MULTIPLE EXCITON GENERATION IN QUANTUM DOTS

Oksengendler B.L., Turaeva N.N., Rashidova S.S.

Institute of Polymer Chemistry and Physics, Uzbekistan Academy of Sciences,
100128 Tashkent, Kadiriy street, 7B, Uzbekistan.
E-mail: nturaeva@hotmail.com

**Abstract**

The statistical theory of multiple exciton generation in quantum dots which is based on the Fermi approach to the problem of multiple particle in nucleon-nucleon collisions is presented. Our estimates of the multiple exciton generation efficiency in different quantum dots, induced by absorption of a single photon, are in a good agreement with the experimental data.

Carrier multiplication (CM) phenomenon in confined nanostructures was predicted by A.Nozik of NREL in 2002 [1]. In 2004, Klimov and co-authors reported that PbSe nanocrystals can respond to absorption of a single photon by producing two electron-hole pairs with 200% efficiency. This carrier multiplication phenomenon, also known as multiple exciton generation (MEG) [2]. Later the effect of MEG has been _experimentally_ confirmed by NREL team [3] by investigating PbSe and PbS quantum dots of 3 different sizes in which increasing of quantum yield up to 300% has been observed. Recently it was revealed by Schaller et al. [4] that MEG is the real phenomenon with generation to up 7 excitons in PbSe QDs ($E_g$=0.3 eV and R=20 nm) at absorption of a single photon with energy of h$\nu$=7.8 $E_g$.

Using carrier multiplication Klimov [4,5] has demonstrated "exotic" non-Poissonian distributions of carrier populations. Different groups have reported recently the formation of multiple excitons based on the femtosecond transient spectroscopic observations. Possible applications of the effect in constructions of on Polymer-QD based solar cells [6] and for low threshold laser design [7,8].

It should be noted that besides practical applications the MEG effect can be used as a testing ground for new kind of nonlinear photo effect. The theoretical analysis of MEG related to two or three excitons generation has been considered in several studies by Schaller R.D. et al. [9] and Elingson et al. [3]. Despite the considerable amount of papers on MEG the problem of theoretical explanation of MEG of higher multiplicity in QDs still remains open. In our opinion, this is caused by the fact that the application of perturbation theory to the charge multiplication effect (e.g. as used in [9]) is complicated for the processes of higher multiplicity. Recently some other approaches to theoretical understanding the mechanism of MEG have been considered (see [9,10]). However the final conclusion is that the mechanism of MEG is not fully revealed yet.

Therefore other mechanisms need to be considered for explaining the highest MEG efficiencies. In addition, the above mentioned non-Poissonian character of exciton multiplicity which has been recently experimentally observed [5] needs also to be explained from the in theoretical viewpoint.

In this paper we present a mechanism of MEG in QDs which can be treated using the combination of statistical theory and on Fermi approach [11] to the multiple particle creation inhigh-energy nucleon-nucleon collisions. The latter was originally used for the analysis of the generation of elementary particles – nucleons and mesons in high-energy nucleon-nucleon collisions. Unlike the perturbation theory used in [9], this approach is based on the strong interaction of correlated electrons with electromagnetic field in a QD. We argue that the Fermi theory seems to be closer to the reality in a high energy range when the number of possible exciton states with given energy is large, and namely this factor sharply increases the probability of statistic equilibrium setting.

In particular, we suppose that in the case of multiple excitons generation the high energy photon is absorbed by a QD and according to statistic laws the absorbed energy is quickly distributed among the various degrees of freedom. Namely, in a QD of $\Omega$ volume $n/2$ excitons are generated where n is the total number of produced particles (electrons and holes). In this case using the Fermi`s formula we can calculate the probability for n-particles generation with a given energy distribution in a small volume of QD. Here we take into account the fact that the total kinetic energy is defined by the difference of photon energy and energy gap multiplied by the number of excitons $\left(T = h\upsilon - \frac{n}{2}E_g\right)$. Then it can be shown that the probability for n-particles generation in the volume of $\Omega$ is proportional to the statistic weight:

$$S(n) = \frac{m^{3n/2}\Omega^n}{2^{3n/2}\pi^{3n/2}\hbar^{3n}} \frac{\left(h\upsilon - \frac{n}{2}\tilde{E}_g\right)^{\frac{3n}{2}-1}}{\left(\frac{3n}{2}-1\right)!}.$$

Here m is the electron mass, $\tilde{E}_g = E_g - \frac{1.78 \times e^2}{\varepsilon R}$ [14], in which the second term describes the electron-hole interaction energy, R is the radius of QD and the number of particles is even number: n= 2,4,6,8,10,12,14.

Then the relative probability for n-particles generation is given by the following expression:

$$W(n) = \frac{S(n)}{\sum_n S(n)}.$$

The quantum efficiency of multiple excitons generation by a single photon can be written as $QE = 100\% \times \langle N_{exc} \rangle$, where the average number of excitons in QD $\langle N_{exc} \rangle$ is calculated using on the following equation:

$$\bar{n} = 2\langle N_{exc} \rangle = \frac{\sum_n n S(n)}{\sum_n S(n)}.$$

Now we consider application of the above results for MEG in different QDs.

1. For the case of PbSe QDs ($E_g$=0.64eV, R=3.9nm, hν=3.63$E_g$) we have $\langle N_{exc}^{theory} \rangle = 2.00$ which is in a good agreement with the experiment ($\langle N_{exc}^{exp} \rangle = 2.00$) [9].

2. For PbSe QDs ($E_g$=0.64eV, R=3.9nm, hν=4.9$E_g$) we have $\langle N_{exc}^{theory} \rangle = 2.85$ and $\langle N_{exc}^{exp} \rangle = 3.25$ [9].

3. For PbSe QDs ($E_g$=0.3eV, R=20nm, hν=7.8$E_g$) we have $\langle N_{exc}^{theory} \rangle = 6.84$ and $\langle N_{exc}^{exp} \rangle = 7.0$ [9].

4. For Si QDs ($E_g$=1.2eV, R=9.5 nm, hν=3.4$E_g$) we have $\langle N_{exc}^{theory} \rangle = 2.62$ and $\langle N_{exc}^{exp} \rangle = 2.6 \pm 0.2$ [10].

On the basis of the above statistic approach, it is possible to explain the non-Poissonian character of excitons multiplicities, observed experimentally in the Ref. [5]. For example, in the case of PbSe QD (R=20nm, hν=7.8$E_g$) we can show that $\langle n \rangle = 13.7$, $\langle n \rangle^2 = 187.1$, $\langle n^2 \rangle = 188$ that is $\overline{n^2} \neq (\bar{n})^2 + \bar{n}$. It means that the condition for Poissonian distribution is not valid for the effect of MEG as it was shown in [5].
Furthermore, we suppose that the microscopic mechanism of the effect is connected to the shaking effect of several electrons caused by the primary exciton at absorption of high energy photon. Such effect is studied in detail by several authors (see, e.g., reviews [12] and references therein)Using the same approach as that for the calculation of multiple ionization of atoms by its shaking in the collisions with fast ions, we can estimate the MEG cross section, i.e. the cross section for generation of n electron-hole pairs induced by one-photon-absorption:

$$\sigma_n = \sigma_1^{ex} I_{n-1}^2 \propto \left(\frac{a}{R}\right)^{\gamma}.$$

Here $\sigma_1^{ex}$ is the cross-section of primary exciton generation, $I_{n-1}$ is the overlapping integral of n-1 electrons wave functions in valence and conductive zones of QD, $\gamma$ is the parameter which proportional to multiplicities of n and $a$ is the atom size. It is important to note that the overlapping integral is different from zero due to strong electron-electron correlation in quantum dots.

**Conclusion**

Thus we have treated multiple electron generation in quantum dots induced by one-photon absorption. To explain the effect at macroscopic level we have proposed developed an approach consisting of the combination of statistics and Fermi approach to multiple particle generation in high energy nucleon-nucleon collisions. Our approach is based on the simple assumption that the probability of n particles generation in the volume QDs is defined by the statistic weight S(n), which depends on many parameters: the size of QD, the photon energy, the gap energy, the exciton bind energy, electron and hole effective masses. The microscopic mechanism of MEG which is based on the electrons shaking theory in atomic collisions is developed. Such a mechanism allows take into account the strong electron-electron correlation in QDs caused by quantum confinement of electronic excitations.

**References**

1. Nozik A. "Quantum dot solar cells". J. Phyica E, v. 14, 115, 2002
2. Schaller R.D., Klimov V.I. "High efficient carrier multiplication in PbSe nanocrystals: Implication for solar energy conservation". Phys. Rev. Lett., v.92, 186601, 2004.
3. Ellingson R.J., Beard M.C., Jonson J. et al. "High efficient multiple excition generation in colloidal PbSe and PbS quantum dots". Nano Lett., v. 5, 865, 2005.
4. Schaller R.D., Petruska M.A., Klimov V.I., "Effect of electronic structure on carrier multiplication efficiency: Comparative study of PbSe and CdSe nanocrystals". Appl. Phys. Lett., v.87, 253102, 2005.
5. Schaller R.D., Klimov V.I. "Non-Poissonian Electron population in Semiconductor Nanocrystals via carrier multiplication". Phys. Rev. Lett., v.96, 097402, 2006.
6. Lewis N.S., Crabtree G.W. "Basic research needs for solar Energy Utiligation". In: "Report of the Basic Energy Science Workshop on Solar Energy Utilization". 18021. Second Printing: U.S. Department of Energy (DOE). Office of Basic Energy Science. Washington, DC. 2005.
7. Klimov V.I., "Nanocrystal Quantum Dots: From fundamental photophysics to multicolor lasing". Los Alamos Science, № 8, 214,2003.
8. Klimov V.I., Mikhailovsky A.A., Hollingsworth A., et al. "Stimulated Emission and Lasing in Nanocrystal Quantum Dots". In: "Quantum Confinement: Nanostructured Materials and Devices". Proc. Electrochem. Soc., v.19, 321, 2001. Ed.by M. Cahay, J. Leburton, D. Lockwood.
9. Schaller R.D., Agranovich V.M., KlimovV.I. "High-efficiency carrier multiplication through direct photogeneration of multi-exciton via virtual single-exciton". Nature Phys. v.1., 189, 2005.
10. Oksengendler B.L., Turaeva N.N., Rashidova S.S. "Statistic Theory of Multiple Exciton generation in Quantum Dots". Appl. Solar Energy. № 3, 36, 2009.
11. Fermi E. "Nuclear Processes at High Energy". Prog. Theor. Phys., v. 5, 570, 1950.
12. Dykhne A M and Yudin D L  Sov. Phys.—Usp. v. **21,** 549 (1978)
13. Matveev V.I., Parilis E.S., Shaking at electronic transitions in atoms, Usp.Phys.Nauk, v.138, 573, 1982.
14. Matveev V I  Phys. Part. Nucl. **26,** 329, 1995